\documentclass[superscriptaddress,twocolumn]{revtex4}
\usepackage{bm}
\usepackage{graphicx}
\usepackage{hyperref}

\begin{document}

\title{Electronic states in cylindrical core-multi-shell nanowire}

\author{A. O. Rudakov}
\affiliation{Institute of Physics and Chemistry, Mordovia State
University, 430005 Saransk, Russia}
\author{I. A. Kokurin}
\email[E-mail:]{kokurinia@math.mrsu.ru} \affiliation{Institute of
Physics and Chemistry, Mordovia State University, 430005 Saransk,
Russia} \affiliation{Ioffe Institute, 194021 St. Petersburg, Russia}
\affiliation{St. Petersburg Electrotechnical University ``LETI'',
197376 St. Petersburg, Russia}

\begin{abstract}
The recent advances in nanowire (NW) growth technology have made
possible the growth of more complex structures such as
core-multi-shell (CMS) NWs. We propose the approach for calculation
of electron subbands in cylindrical CMS NWs within the simple
effective mass approximation. Numerical results are presented for
${\rm GaAs/Al_{0.3}Ga_{0.7}As}$ radial heterostructure with
AlGaAs-core and 4 alternate GaAs and AlGaAs shells. The influence of
an effective mass difference in heterolayers is discussed.
\end{abstract}

\date{\today}

\maketitle

\section{Introduction}

The recent progress in nanowire (NW) growth technology gives an
opportunity to made the NW-based complex structures with so-called
axial \cite{Nylund2016} and radial (core-shell) heterostructuring.
Now there is a possibility to grow a large number of shells (see,
for instance, reviews \cite{Lieber2007,Royo2017} and references
therein). Such structures known as core-multi-shell (CMS) NWs
attract attention of researchers due to interesting properties and
possible applications.

Due to a special geometry and the possibility to govern carrier
states by means of external fields (electric, magnetic or
deformation) the devices based on NW-structures \cite{Hayden2008}
are very attractive for modern electronics and photonics. For
example, there are lasers \cite{Saxena2016,Stettner2016} and
light-emitting diodes \cite{Tomioka2010} with CMS NW as a work item.

The carrier mobility can increase in CMS structures \cite{Funk2013}
comparative to continuous NW, that means the possibility of
conductance quantization in such structures. Thus, CMS NWs are good
candidates for utilizing as the working part of field effect
transistors including spin ones.

The realization of multiple quantum wells in CMS NWs can lead to the
spatial separation between the electron and hole, that allows one to
control the lifetime of indirect excitons \cite{Butov2017}, which
can travel over large distances before recombination, and cool down
close to the lattice temperature and below the temperature of
quantum degeneracy.

Thus, there is a need to know electron and hole subband spectrum in
such structures. Here we develop a simple approach to find
electronic states in ${\rm III-V}$ CMS NW with zinc-blende crystal
lattice within single band effective mass approximation (EMA).

\section{Hamiltonian and numerical diagonalization}

Now we propose the approach for calculation of electron subband
spectrum of CMS NWs. We use a single-band envelope function
approximation (EFA) to find electronic states in cylindrical CMS NW
with zinc-blende crystal structure (starting from bulk
$\Gamma_6$-band states with scalar effective mass).

In cylindrical NW there are translational and rotational invariance,
that means the following form of envelope wavefunction

\begin{equation}
\label{gen_func}
\Psi_{mnk}(r,\varphi,z)=R_{mn}(r)\frac{1}{\sqrt{2\pi}}e^{im\varphi}\frac{1}{\sqrt
L}e^{ikz},
\end{equation}
where $k$ is the longitudinal momentum, $m=0,\pm 1, \pm 2,...$ $n=1,
2,3, ...$. The energy spectrum of one-dimensional (1D) subbands is
given by

\begin{equation}
\label{gen_energy} E_{mn}(k)=\epsilon_{mn}+\frac{\hbar^2k^2}{2m^*},
\end{equation}
where $m^*$ is the scalar conduction band effective mass, and
$\epsilon_{mn}$ is the energy of 1D-subband bottom.

For the case of uniform cylindrical NW of radius $R$ applying zero
boundary conditions (hard-wall potential) we have for $R_{mn}$ in
Eq.~(\ref{gen_func}) and $\epsilon_{mn}$ in Eq.~(\ref{gen_energy})
well known result
\begin{equation}
\label{basis}
R^0_{mn}(r)=\frac{\sqrt
2}{R|J_{|m|+1}(j_{mn})|}J_{|m|}\left(j_{mn}\frac{r}{R}\right),
\end{equation}
and
\begin{equation}
\label{energy0}
\epsilon^0_{mn}=\frac{\hbar^2j_{mn}^2}{2m^*R^2}.
\end{equation}

Here $J_m(x)$ is the first kind Bessel function of order $m$, and
$j_{mn}$ is the $n$-th root of function $J_m$, i.e. $J_m(j_{mn})=0$,
with $n=1,2,3,...$ .

\begin{figure}
\includegraphics[width=\linewidth]{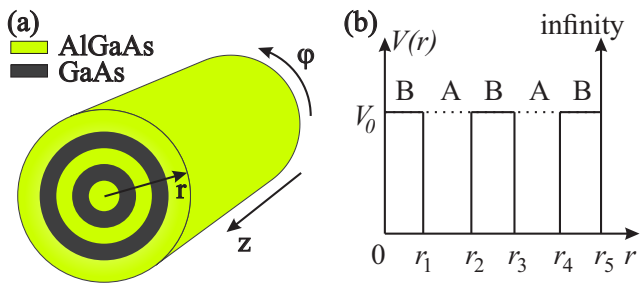}
\caption{\label{fig01} (a) The sketch of CMS NW. The case of
two-well-three-barrier structure is depicted, that can be realized
by means of usual hetero-pair GaAs-Al$_x$Ga$_{1-x}$As. (b) An
effective confining potential which replicates the conduction band
profile for the structure depicted in (a). Character sizes (radii)
discussed in text are depicted. The conduction band offset is equal
to $V_0$.}
\end{figure}

Now consider the spectral problem for CMS NW. The one-electron
Hamiltonian within single-band EMA is $H_0+V$, with $H_0$ being the
Hamiltonian of uniform NW, that corresponds to wavefunction
(\ref{basis}), and $V(r)$ is the potential giving the difference
between NW and CMS NW. The conduction band offset serves as
effective radial potential
\begin{equation}
\label{potential} V(r)=\left\{
\begin{array}{ll}
0 & \textrm{in wells}\\
V_0 & \textrm{in barriers}
\end{array} \right.
\end{equation}
is depicted in Fig.~\ref{fig01}b. In reality such a structure must
have GaAs outer capping layer in order to avoid Al oxidation and the
following degradation of the structure. We suppose that such a
capping layer usually thin and we do not make a mistake in wave
function behavior neglecting them.

Translation and rotational invariance are conserved in this case and
wave function will have form (\ref{gen_func}) but with $R_{mn}$
different from $R^0_{mn}$ of Eq.~(\ref{basis}). However, we can look
for wave function as a series on basis functions of
Eq.~(\ref{basis}).

\begin{equation}
\label{series} R_{mn}(r)=\sum_lC^{mn}_lR^0_{ml}(r).
\end{equation}

We can find the Hamiltonian $H_0+V(r)$ matrix in basis
(\ref{basis}). It will contain $\epsilon^0_{ml}$ in diagonal and
matrix elements of (\ref{potential})
\begin{equation}
V^m_{l'l}=V_0\sum\limits_{i=0}^2\int\limits_{r_{2i}}^{r_{2i+1}}drrR^0_{ml'}(r)R^0_{ml}(r)
\end{equation}
in all positions. Where $r_i$ ($i=1-5$) is the radius of $i$-th
radial heterointerface, and $r_0=0$, $r_5\equiv R$. In chosen basis
the matrix elements $V^m_{l'l}$ can be found analytically
\cite{Prudnikov1986}, but we do not write them here due to their
cumbersome form.

\begin{figure}
\includegraphics[width=0.85\linewidth]{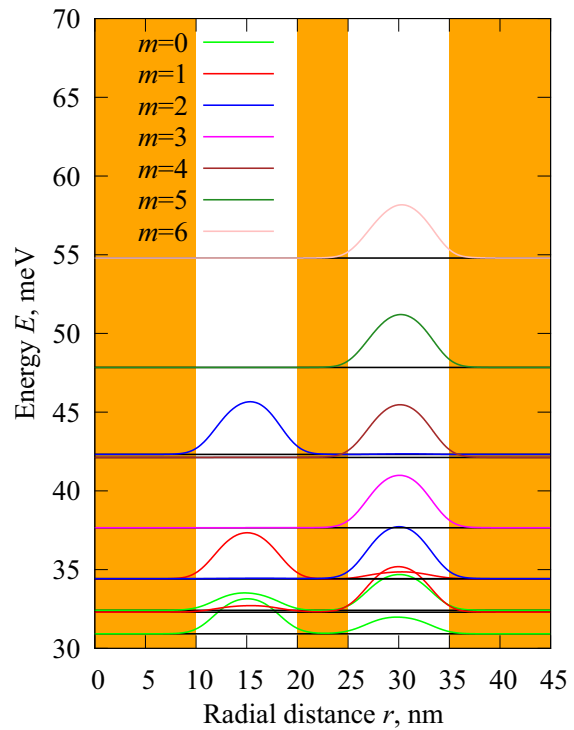}
\caption{\label{fig02} The CMS NW subband energies $\epsilon_{mn}$
and corresponding radial probability densities $r|\Psi|^2$. The
first 10 subband energies presented for the structure with
$r_i=0,10,20,25,35,45$ nm. $V_0=230$ meV, $m^*=0.067m_0$. The areas
corresponding to wells and barriers are transparent and shaded,
respectively.}
\end{figure}

The numerical diagonalization was performed for CMS NW with $45$ nm
radius and the widths of well and barrier regions depicted in
Fig.~\ref{fig02} (transparent and shaded areas, respectively). The
finite barriers are of 230 meV height, that approximately
corresponds to conduction band offset at heterointerface ${\rm
GaAs/Al_xGa_{1-x}As}$ with $x=0.3$ \cite{Vurgaftman2001}. Effective
mass was chosen as in GaAs, $m^*=0.067m_0$. We used a finite
dimension Hamiltonian of $40\times 40$ dimension, that gives the
perfect precision (better than 0.1\%) for first 10 subbands in each
block with fixed $m$. Besides 1D-subband bottoms the coefficients
$C^{mn}_l$ found giving coordinate dependence of wavefunction in
accordance with Eqs.~(\ref{gen_func}),(\ref{series}),(\ref{basis}).

\section{Discussion}

The results of numerical diagonalization for first ten 1D-subband
bottoms are depicted in Fig.~\ref{fig02} as well as corresponding
probability densities $r|\Psi|^2$ found from
Eqs.~(\ref{series}),(\ref{basis}). One can see expected behavior of
wave function: for low-lying states belonging to the family with the
same $m$ the ground and first excited state wave functions are
predominantly localized in different wells. For high-lying subbands
wave functions will have nodes inside each well region. As in
uniform NW the subband ground state corresponding to higher
$|m|$-value has a higher energy. However, relative position of
subband with $\epsilon_{|m|+1,1}$ and subband with
$\epsilon_{|m|,2}$ crucially depends on $V_0$ and relation between
$r_i$.

Till now we supposed the equal effective masses both in wells and in
barriers. In real structures it is necessary also to take into
account a difference of effective masses in different cylindrical
layers. In ${\rm Al_xGa_{1-x}As}$ with $x=0.3$ the effective mass
value is $0.092m_0$ \cite{Vurgaftman2001}. This difference does not
distort the translational and rotational symmetry of the structure.
However, in this case the motion along and across CMS NW is not
formally separated. This means that Hamiltonian matrix remaining
diagonal in $m$ and $k$ will parametrically depend on the
longitudinal momentum $k$. In order to include difference in masses
into our scheme we have to use instead of Eq.~(\ref{potential}) the
following operator

\begin{eqnarray}
\nonumber
\widetilde{V}(r)=\frac{\hbar^2}{2}\left(\frac{1}{m_A}-\frac{1}{m_B}\right)\sum_{j=1}^4(-1)^j\delta(r-r_j)\frac{\partial}{\partial
r}\\
+\left\{
\begin{array}{ll}
0 & \textrm{in wells}\\
-\left(1-\frac{m_A}{m_B}\right)H_0+V_0 & \textrm{in barriers}
\end{array} \right.,
\end{eqnarray}
where $m_A$ and $m_B$ are effective masses in wells and barriers,
respectively. The second term takes into account the difference in
the kinetic energy that is due to the difference of effective masses
in layers. The first term ensures the Hermiticity of the operator,
and it arises from the standard form of the kinetic energy operator
$-\frac{\hbar^2}{2}\nabla\frac{1}{m^*({\bf r})}\nabla$ in the
systems with spatially inhomogeneous mass (the step-like dependence
of $m^*(r)$ leads to $\delta$-function contribution). It is obvious,
that at $m_A=m_B$ we have the same result as before. Due to the
translational invariance along the NW axis one can simply replace in
$H_0$ the longitudinal momentum operator
$p_z=-i\hbar\partial/\partial z$ by its eigenvalue $\hbar k$.

The spectral problem in this case can be solved in the same manner
but in order to find $E_{ml}(k)$ one have to diagonalize total
Hamiltonian at each $k$ value, giving $\epsilon_{ml}$ to be the
function of $k$. This leads to change in subband bottom energy and
renormalization of effective masses in different subbands and
nonparabolicity as well due to different penetration of wave
function into the barriers. These results will be published
elsewhere.

\section{Conclusion}

We proposed the approach for calculation of conduction band states
in CMS NW. The numerical results for five radial-layer ${\rm
GaAs/Al_{0.3}Ga_{0.7}As}$ CMS NW of 45 nm radius are presented. This
approach can be easily generalized to describe hole states in
complex valence band or multiband Hamiltonian in CMS NW of
narrow-gap semiconductor.

\end{document}